# Shot Noise of Non-Interacting Composite Fermions

R. de Picciotto

*Braun Center for Submicron Research, Dep. of Condensed Matter Physics, Weizmann Institute of Science, Rehovot 76100, Israel*

A simple equivalent circuit, which describes transport properties of a Two Dimensional Electron Gas in the Fractional Quantum Hall regime is presented. The physical justifications for this equivalent circuit are discussed in the frame work of the non-interacting Composite Fermions model. Quantum Shot Noise at an arbitrary filling factor and temperature is readily calculated.

Recently two experimental works confirmed the existence of fractionally charged ($e/3$) 'Laughlin quasiparticles' by utilizing *quantum shot noise* measurements in the Fractional Quantum Hall (FQH) regime [1, 2]. Shot noise results from the discrete nature of the current carrying charges; hence is proportional to their charge and to the average current *I*. Theoretical predictions for the expected shot noise in the FQH regime were made within the framework of the chiral Luttinger Liquid model [3, 4, 5, 6]. Both the I-V characteristics and the zero frequency spectral density of current fluctuations, *S*, were calculated for the case of a partly reflecting impurity embedded in a Two Dimensional Electron Gas (2DEG) subjected to a strong perpendicular magnetic field at a Filling Factor (FF) ν=1/3. Zero temperature analytical results were obtained [3] in the two limiting cases of weak and strong reflections by the impurity. In the weak reflection limit the reflected current is expected to be composed of quasi particles; hence the noise is proportional to the reflected current, $I_r$, and to the quasi particle charge, $e^* = e/3$, namely, $S = 2I_r e^*$. The above-mentioned experiments [1, 2] were performed in this regime and confirmed this prediction. On the other hand, in the strong reflection limit, charges are rarely transmitted through a thick and high barrier region. The current is thus expected to be dominated by electrons and shot noise is expected to be proportional to the transmitted current, $I_t$, and to the electronic charge, *e*, namely, $S = 2I_t e$. Numerical calculations, performed for an arbitrary reflection coefficient [4] and temperature [6] show that the noise changes its characteristics from thermal (Johnson-Nyquist) noise in equilibrium to shot noise at large bias. Shot noise, calculated at sufficiently large applied voltages, reflects the inherent dependence of the reflection coefficient on the applied voltage and the non-trivial dependence of the apparent quasiparticle charge on the reflection coefficient.

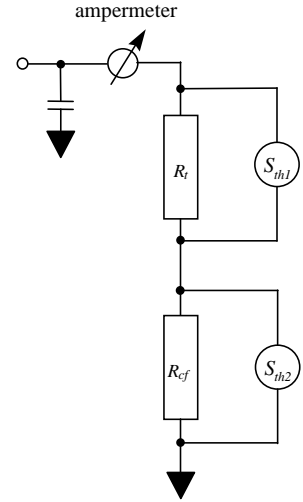

**FIG. 1.** The equivalent circuit describing the resistor due to attachment of two flux quanta to each electron, $R_t$, in series with a Hall resistor representing the residual field $\Delta B$, $R_{cf}$. A thermal noise source is added to each resistor. An ampermeter measures current fluctuations due to a finite temperature *T*.

Another approach, frequently used in the context of the FQH effect, is the Composite Fermion theory [7]. Here a singular (Chern-Simon) gauge transformation is used in order to transform the problem of strongly interacting electrons into another problem of new 'composite particles' in the presence of a reduced magnetic field. The hope here is that the reduced degeneracy may suppress the role of interactions among these new quasiparticles. Within a mean field approximation these new composite particles can be viewed as being composed of the original electrons, each with an even number of attached flux tubes - each tube contains one flux quantum, $\phi_0 = h/e$ - that are derived from the original magnetic field. The composites thus move about in the presence of a smaller, residual, magnetic field. Indeed, the fractional FF's at which the FQH effect is observed correspond to integer FF's of the composites, namely, to the Integer Quantum Hall (IQH) effect [8] of these composites in the presence of the residual field.

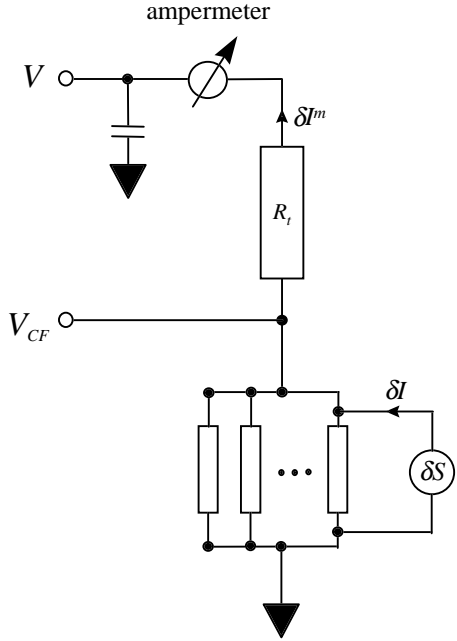

**FIG. 2.** The equivalent circuit of an embedded QPC which controls the conductance of the CF channels. A current noise source is added to the parallel resistors, representing the excess noise. An ampermeter measures current fluctuations due to an applied constant voltage $V$.

Our aim here is to calculate the expected shot noise within the mean field approximation of the non-interacting CF model. The DC properties of this model will be briefly reviewed bellow and the noise properties that are suggested will follow.

For simplicity, let us consider the attachment of two flux quanta to each electron and assume that the external magnetic field, $B$, is larger than $B_{1/2} = 2\phi_0 n_s$ ($n_s$ being the 2DEG areal density) by $\Delta B$. The FF of the CF, $p = \phi_0 n_s / \Delta B$, in that case is related to the FF of the electrons, $\nu = \phi_0 n_s / B$, via [7]:

$$\frac{1}{\nu} = \frac{1}{½} + \frac{1}{p} \Rightarrow \nu = \frac{p}{2p+1} . \quad (1)$$

The resistivity tensor of the CF, $\rho_{cf}$, is related to that of the electrons, $\rho$, by [7]:

$$\rho = \rho_{cf} + 2\frac{h}{e^2}\begin{bmatrix} 0 & 1 \\ -1 & 0 \end{bmatrix}. \quad (2)$$

At certain magnetic field values, corresponding to Hall resistance plateaus in the FQH regime, $\rho_{cf}$ in Eq. (2) is off diagonal, implying that the total Hall voltage equals the sum of two parts and the tensor relation in Eq. (2) can be reduced to a scalar relation of the form:

$$R = R_{cf} + R_t = \frac{h}{e^2}\left(\frac{1}{p} + 2\right), \quad (3)$$

with the $R$'s being the off diagonal components. This relation is depicted in the trivial equivalent circuit in Fig. 1, with the two resistors connected in series as in Eq. (3).

Consider now a Quantum Point Contact (QPC) [9, 10] embedded in a 2DEG in the FQH regime in the bulk. The QPC is described by a set of $p$ transmission coefficients, $\{t_i\}$, for the various CF Landau levels through the QPC [11, 12] (assuming no mode mixing). The transmission coefficients are controlled by an applied gate voltage. This is represented in the equivalent circuit by replacing $R_{cf}$ by $p$ parallel resistors each representing a single CF Landau level (Fig. 2). Each resistor's conductance equals $g_0 \cdot t_i$, where $t_i$ is the transmission coefficient of the $i$'th CF Landau level through the QPC. The conductance of the whole system:

$$g = g_0\left[2 + \left(\sum_{i=1}^{p} t_i\right)^{-1}\right]^{-1}, \quad (4)$$

is expected to exhibit plateaus at conductance's $g_j = g_0 \cdot j/(2j+1)$ where $j \leq p$ equals an integer corresponding to the number of fully transmitted CF Landau levels, in agreement with experimental observations.

We now turn to the analysis of shot noise properties of this equivalent circuit. We make use of the general formula for zero frequency spectral density of quantum shot noise, applicable to non-interacting Fermions [13]; $S = S_{th} + \delta S$, where $S_{th} = 4k_B T g$ is the thermal noise of the sample in equilibrium at a temperature $T$, and the so called excess noise, $\delta S$, is given by;

$$\delta S(V) = 2g_0 \sum_i t_i (1-t_i) \left[ eV \coth\left(\frac{eV}{2k_B T}\right) - 2k_B T \right], \quad (5)$$

with $t_i$ the transmission probability of each channel and $V$ the applied voltage across the QPC. In equilibrium ($V=0$) both resistors in Fig. 1 generate thermal noise. These two (uncorrelated) noise sources (denoted as $S_{th1}$ and $S_{th2}$) result in a net measured spectral density of current fluctuations in the ampermeter, $S_{th}^m = 4k_B T g$, where $g$ is given by Eq. (4); as expected from the fluctuation-dissipation theorem. For a finite $V$, $\delta S > 0$, as represented in Fig. 2 by the ideal current source connected in parallel with the resistors representing $R_{cf}$. The crucial point to note here is that for fully transmitting QPC no excess noise is generated. Namely, the upper resistor, $R_t$, which merely represents a mathematical transformation does not generate excess noise (neglecting current related heating effects such as the ones discussed in [14]).

Assuming the CF's are non-interacting, the excess noise generated due to partitioning of CF Landau levels is given by Eq. (5) with $V$ replaced by, $V_{cf} = V R_{cf} / (R_{cf} + R_t)$, the voltage drop on the CF channels (see Fig. 2). Note that the full charge $e$ as well as the quantum conductance $g_0$ are being used for the CF's since the effect of the two fluxes attached to each electron is embedded in $R_t$. The relation of this Fermionic shot noise to the statistics of quasiparticles, which is believed to be fractional [15], will be discussed later on. Note that the current fluctuations measured in the ampermeter are even smaller than $\delta S(V_{cf})$ since the current fluctuations divide between the two branches of the circuit, namely, $R_{cf}$ and $R_t$ (see Fig. 2). Hence the measured noise, $\delta S^m$, is expected to be:

$$\delta S^m = \delta S(V_{cf}) \left[ \frac{R_{cf}}{R_{cf} + R_t} \right]^2 . \quad (6)$$

The next step is to associate this expected noise suppression with a fractional charge. In order to do that, we view the conductance plateaus (suggested by Eq. (4)) as accumulative contributions of subsequent channels. Contrary to the non-interacting Fermions case, where each channel contributes the same amount, $g_0$, to the total conductance, hence the current divides equally among the available channels, here each channel contributes a different portion to the conductance, $\delta g_j$, hence carries a different portion of the total current, as shown in Fig. 3. We define the transmission coefficients of these channels, $\tau_j$, by;

$$g = \sum_{j=1}^{p} \delta g_j \tau_j , \quad (7)$$

where $\delta g_j = g_j - g_{j-1}$. This in fact implies that we construct another equivalent circuit, shown in the inset of Fig. 3, which resembles the equivalent circuit of non-interacting Fermions (Fig. 2 without the upper resistor - $R_t$).

For $(j-1)$ fully transmitted CF channels and a $j$'th partially transmitted channel, one obtains the following relation between $t_j$ in Eq. (4) and $\tau_j$ in Eq. (7):

$$\tau_j = t_j \frac{2j+1}{2j + 2t_j - 1} . \quad (8)$$

Using Eqs. (5-8), the expected QSN, $\delta S_I^m$, can be rewritten in the form:

$$\delta S_I^m = 2\delta g_j \tau_j (1-\tau_j) \left[ qV \coth\left(\frac{qV}{2k_B T}\right) - 2k_B T \right], \quad (9)$$

with an effective charge, $q$, which is $\tau_j$ dependent,

$$q = \frac{e}{2j-1}\left[1 - \frac{2\tau_j}{(2j+1)}\right], \quad (10)$$

making Eq. (9) similar to Eq. (5). Namely this excess noise is *equivalent* to partition noise generated due to the partial transmission, $\tau_j$, of the impinging current of the *j'th* channel, $V\delta g_j$, with *the current carrying charge being equal to q*. The charge, $q$, which follows from Eq. (10), varies linearly with the transmission coefficient, $\tau_j$, from $e/(2j+1)$ in the weak back scattering limit ($\tau_j=1$) to $e/(2j-1)$ in the strong back scattering limit ($\tau_j=0$). Hence for example, for $\nu=1/3$ ($j=1$) the charge in the limit of weak back scattering is $e/3$ while in the limit of strong back scattering the charge is $e$. In these two extreme cases, the same conclusion was reached by calculations based on a chiral Luttinger liquid model [3]. Note that the interacting CF model (when considered beyond the mean field approximation) maps onto the Luttinger liquid model [16, 17].

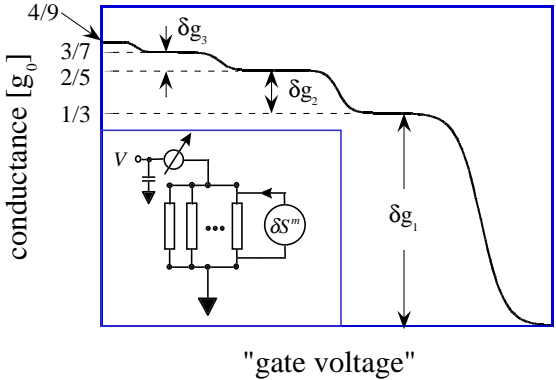

**FIG. 3.** Conductance plateaus, expected from Eq. (4), corresponding to sequential reflection of CF Landau levels. A FF of 4/9 in the bulk is assumed. The contributions of the first three channels are shown. **Inset:** The equivalent circuit which describes *p* accumulative contributions, $\delta g_j$, to the total conductance.

An intuitive way to understand the result of Eq. (10) is to note that this reduced charge equals the net charge which is transferred through the QPC when a CF is transmitted. The flux tubes motion induces currents which carry charge away from the QPC (eventually into Ohmic contacts), resulting in a net charge, $q$, being transferred;

$$q = e - \int dt\, g\, \frac{d\Phi}{dt} = e - 2\phi_0 g = \frac{e}{2j-1}\left[1 - \frac{2\tau_j}{2j+1}\right].$$

Such treatment can be readily generalized to other CF type transformations, e.g. near FF's, $\nu=1/m$ (where *m* is an even integer), by attaching $m=4,6,...$ flux quanta to each electron. It can also be extended to both positive and negative magnetic field deviations $\Delta B$ (the sign of the charge is determined by that of $\Delta B$, however, as is evident from Eq. (5) the excess noise does not depend on this sign). The behavior of quantum shot noise can thus be predicted analytically for any fractional FF and temperature.

The reduced charge of Laughlin's quasiparticles might be related to their fractional statistics [15]. It is thus interesting to compare the result presented above to the statistical properties of Anyons. We adopt an approach put forward by Wu [18] in which the statistical weight (i.e. the number of distinct configurations of *N* particles in *G* levels), $w$, is given by a binomial coefficient of the form:

$$w = \binom{G+(1-\alpha)(N-1)}{N}, \quad (11)$$

where the parameter $0<\alpha<1$ governs the statistics. Fermi-Dirac and Bose-Einstien statistics are recovered for $\alpha=1$ and $\alpha=0$ respectively. We assume here quasiparticles of one species only which correspond to simple quantum Hall fractions with a numerator equal to 1 (for details see [18]). The average occupation of a state with energy $\varepsilon$, $\bar{n}_\varepsilon$, deuced from this statistical weight has the property; $0<\bar{n}_\varepsilon<1/\alpha$, *i.e.* each state can occupy $1/\alpha$ quasiparticles. A simple calculation of the fluctuations of this quantity yields[19]:

$$\overline{\delta n_\varepsilon^2} = \bar{n}_\varepsilon(1-\alpha\bar{n}_\varepsilon)[1+(1-\alpha)\bar{n}_\varepsilon]. \quad (12)$$

Assuming that these statistical weights describe the stochastic process of charge transfer at the QPC, that is, $w$ is the number of distinct configurations of *N* impinging quasiparticles on one edge of the sample in $G$ states on the opposite edge, one may calculate shot noise at zero temperature as;

$$S = 2\frac{\overline{\delta Q_{\Delta t}^2}}{\Delta t} = 2\frac{(e^*)^2 N_{\Delta t}}{\Delta t}\overline{\delta n^2} \ , \quad (13)$$

here $Q_{\Delta t}$ is the charge transferred during a time interval $\Delta t$, $N_{\Delta t}$ equals the number of impinging quasiparticles within this time interval and $e^* = \alpha e$ is the charge of each quasiparticle. The average reflected current is given by;

$$I_r = \frac{e^* N_{\Delta t}}{\Delta t}\overline{n} \equiv V\delta g \cdot r \ , \quad (14)$$

thus $r = \alpha \overline{n}$. Combining Eqs. (12-14) yields:

$$S = 2e^*V\delta g \cdot r(1-r)(1+(\alpha^{-1}-1)r) \ . \quad (15)$$

This result coincides with Eq. (9) for zero temperature, a single partially transmitted CF channel ($j$=1) and $m$ attached flux tubes per electron with $\alpha$=1/(m+1) and $r$=1-$\tau$, as expected. Our result for the QSN in the FQH regime supports thus the particular choice of the statistical weights in Ref. [18].

The author would like to acknowledge fruitful discussions with N. Q. Balaban, A. Finkelstein, M. Heiblum, Y. Imry, S. Levit, F. von Oppen, D. Orgad, M. Reznikov, A. Schwimmer, A. Stern and A. Yacoby. The work was partly funded by a grant from the Israeli Science Foundation and by the "Eshkol" scholarship from the Israeli Ministry of Science.